\documentclass[]{aastex631}

\usepackage{amsmath}	% Advanced maths commands
\usepackage{amssymb}	% Extra maths symbols

\DeclareRobustCommand{\ion}[2]{%
\relax\ifmmode
\ifx\testbx\f@series
{\mathbf{#1\,\mathsc{#2}}}\else
{\mathrm{#1\,\mathsc{#2}}}\fi
\else\textup{#1\,{\mdseries\textsc{#2}}}%
\fi}

\usepackage{graphicx}	% Including figure files
\usepackage{subfigure}

\begin{document}

% \title{Analysis of Cross Matching from the Verry Large Array Sky Survey and the Hyper Suprime Cam Catalog\\[0.5em] \large University of Wisconsin-Madison Cosmology group}

\title{Predicting the Number of Radio Sources Seen by Both VLASS and LSST}

\author{Alex Tellez}
\affil{Department of Physics, University of Wisconsin, Madison, 1150 University Avenue, Madison, WI 53706, USA}

\author[0000-0003-1432-253X]{Yjan Gordon}
\affil{Department of Physics, University of Wisconsin, Madison, 1150 University Avenue, Madison, WI 53706, USA}

% \collaboration{20}{(AAS Journals Data Editors)}

\author[0000-0001-8156-0429]{Keith Bechtol}
\affil{Department of Physics, University of Wisconsin, Madison, 1150 University Avenue, Madison, WI 53706, USA}

\begin{abstract}
Radio surveys typically sample extragalactic sources in higher redshift regimes than is typical for optical surveys, resulting in many radio sources not having a detected optical counterpart.
Over the next decade the Legacy Survey of Space and Time (LSST) will be performing the deepest ($i < 26.4\,$mag) wide-area optical survey to date increasing the fraction of radio sources for which we have optical data.
In this Research Note we use the Hyper Suprime-Cam survey to analyse how the fraction of radio sources in the Very Large Array Sky Survey (VLASS) with optical detections varies as a function of $i$-band magnitude and extrapolate to predict the number of optical counterparts we expect LSST to detect.
Assuming a final VLASS point source depth of $S_{3\,\text{GHz}}\lesssim350\,\mu$Jy, we expect LSST to identify optical counterparts to $\sim 10^6$ radio sources in VLASS.
\end{abstract}

\keywords{
\href{http://astrothesaurus.org/uat/1338}{Radio Astronomy (1338)},
\href{http://astrothesaurus.org/uat/1776}{Optical Astronomy (1776)},
\href{http://astrothesaurus.org/uat/1464}{Sky Surveys (1464)},
\href{http://astrothesaurus.org/uat/1167}{Optical Identification (1167)},
\href{http://astrothesaurus.org/uat/506}{Extragalactic Astronomy (506)}}


%%%word limits
%%%1500 words max - includes [this manuscript]:
%%% title [12]
%%% abstract (150 max) [133]
%%% body [948]
%%% headers [8]
%%% captions [82]
%%% references [164]
%%%
%%% TOTAL = 

%%%%%%%%%%%%%%%%%%%%%%%%%%%%%%%%%%%%%
\section{Introduction} 
\label{sec:intro}

Wide-area radio surveys can be used to study phenomena such as active galactic nuclei (AGN) and star-formation in galaxies.
% Obtaining physically meaningful quantities for radio sources requires distances to the galaxy to be known, even for basic measurements (e.g., converting a flux density to a luminosity), and these distance measurements usually require optical information (typically for redshifts).
% Consequently, optical counterparts are essential in order to effectively study radio sources.
Distance measurements (usually redshifts) needed for physical inferences about radio sources, typically require identification of optical counterparts.
% Unfortunately many counterparts to radio sources remain undetected as current optical surveys lack the sensitivity needed to detect them.
Unfortunately current optical surveys are typically too shallow to identify many counterparts to radio sources.
The Vera C. Rubin Legacy Survey of Space and Time \citep[LSST,][]{Ivezic2019} will conduct a deep ($i\lesssim 26.4\,$mag) and wide-area ($\sim 18,000\,\text{deg}^{2}$) optical survey over the next 10 years \citep{Bianco2022}, improving our ability to detect optical counterparts to radio sources. 
In this Research Note, we use archival survey data to predict the number of optical counterparts that LSST will detect for radio sources identified by the Very Large Array Sky Survey \citep[VLASS,][]{Lacy2020}.

%%%%%%%%%%%%%%%%%%%%%%%%%%%%%%%%%%%%%
\vspace{1em}
\section{Information from Existing Data} 
\label{sec:data}

VLASS is surveying $\sim 80\,\%$ of the sky at $\nu\sim3\,$GHz over three epochs, with typical rms noise of $120\,\mu\text{Jy}\,\text{beam}^{-1}$ per epoch.
% With the highest angular resolution ($2''.5$) of any wide-area radio survey to date, VLASS is well suited to making robust multiwavelength cross identifications \citep{VillarrealHernandez2018}.
The $2''.5$ angular resolution of the survey makes VLASS well suited for robust multiwavelength cross-identifications \citep{VillarrealHernandez2018}.
% To predict the expected number of LSST counterparts to VLASS sources we first need to determine how the fraction of radio sources with a detected optical counterpart varies with optical depth and then extrapolate that to the depth of LSST.
To predict the expected number of LSST counterparts to VLASS sources we first need to determine how the fraction of radio sources with a detected optical counterpart varies with optical depth before extrapolating to the depth of LSST.
% Doing so requires a deep optical catalog with which to match to VLASS.
This requires a deep optical catalog with which to match to VLASS.
% For this we use  the third data release (DR3) from the Hyper Suprime-Cam wide survey \citep[HSC,][]{Aihara2022}, the deepest wide-area optical survey prior to LSST, having a target depth of $i< 26.2\,$mag over $1,400\,\text{deg}^{2}$ and lying entirely within the VLASS footprint.
We use  the third data release (DR3) from the Hyper Suprime-Cam wide survey \citep[HSC,][]{Aihara2022}, which has a target depth of $i< 26.2\,$mag over $1,400\,\text{deg}^{2}$ and lies entirely within the VLASS footprint.

To identify radio sources with an optical counterpart, we select reliable detections from the VLASS epoch 1 catalog \citep{Gordon2021}, specifically those satisfying \texttt{S\_Code} $\neq$ `E', \texttt{Duplicate\_flag} $<2$ and \text{\texttt{Quality\_flag} $==(0|4)$}, matching these to the nearest HSC source.
% This VLASS catalog is based on `limited quality' \textit{Quick Look} images, and as a result is only complete to a point source depth of $3\,$mJy \citep{Gordon2021}. 
The number of spurious matches resulting from random spatial coincidence is estimated by $B = 2\,\pi\, r\,N\,\rho_{0}\,dr$ \citep{Galvin2020}, where $B$ is the number of false positive matches, $r$ is the angular separation between the optical and radio sources, $N$ is the number of radio sources with a match, and $\rho_{0}$ is the source density of the optical survey ($\sim 270,000\,\text{deg}^{-2}$ for HSC).
In Figure \ref{figure}a we show the distribution of angular separations for our VLASS/HSC matches and the expected number of contaminants.
We verify $B(r)$ is a reliable estimate by repeating out HSC cross match with $N$ random sky coordinates, and the distributions of separations are shown to be consistent in Figure \ref{figure}a.
Knowing the number of false positive matches as a function of match separation, the probability that a match is reliable, $P_{\text{real}}(r)$, can be estimated using $1-B(r)/N(r)$.
The astrometric accuracy of the current VLASS data is typically $\approx 0''.5$ \citep{Gordon2021}, and in the upper panel of \text{Figure \ref{figure}a} we see that matches with $r<0''.5$ have $P > 0.8$, with their reliability falling off quickly at larger separations.
We therefore only consider VLASS/HSC pairings with $r<0''.5$ as matches for the rest of this analysis.

%%%%%%%%%%%%%%%%%%%%%%%%%%%%%%%%%%%%%
\vspace{1em}
\section{Predictions for LSST} 
\label{sec:predictions}

\begin{figure}
    \centering
    \subfigure[]{\includegraphics[width=0.48\textwidth]{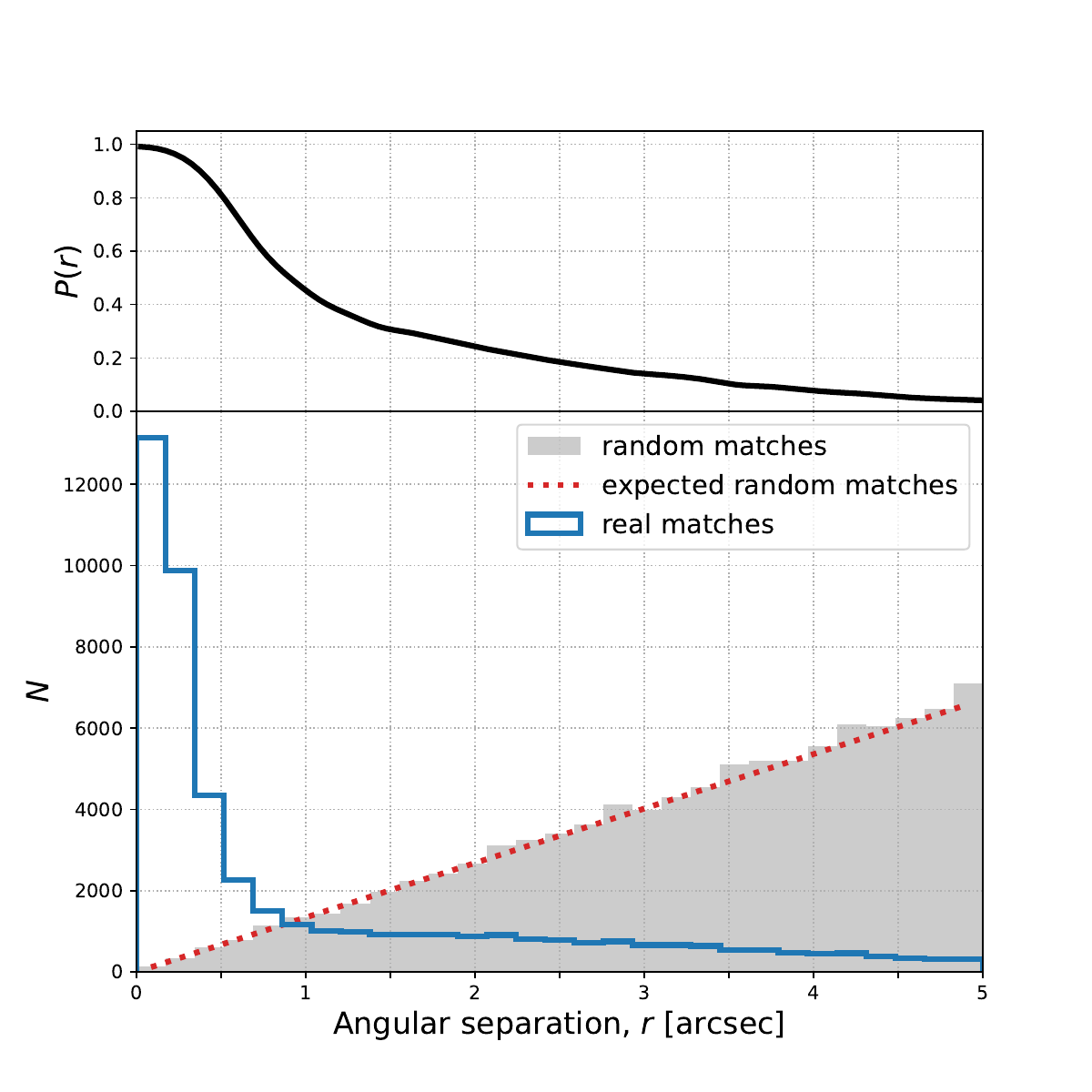}}
    \subfigure[]{\includegraphics[width=0.48\textwidth]{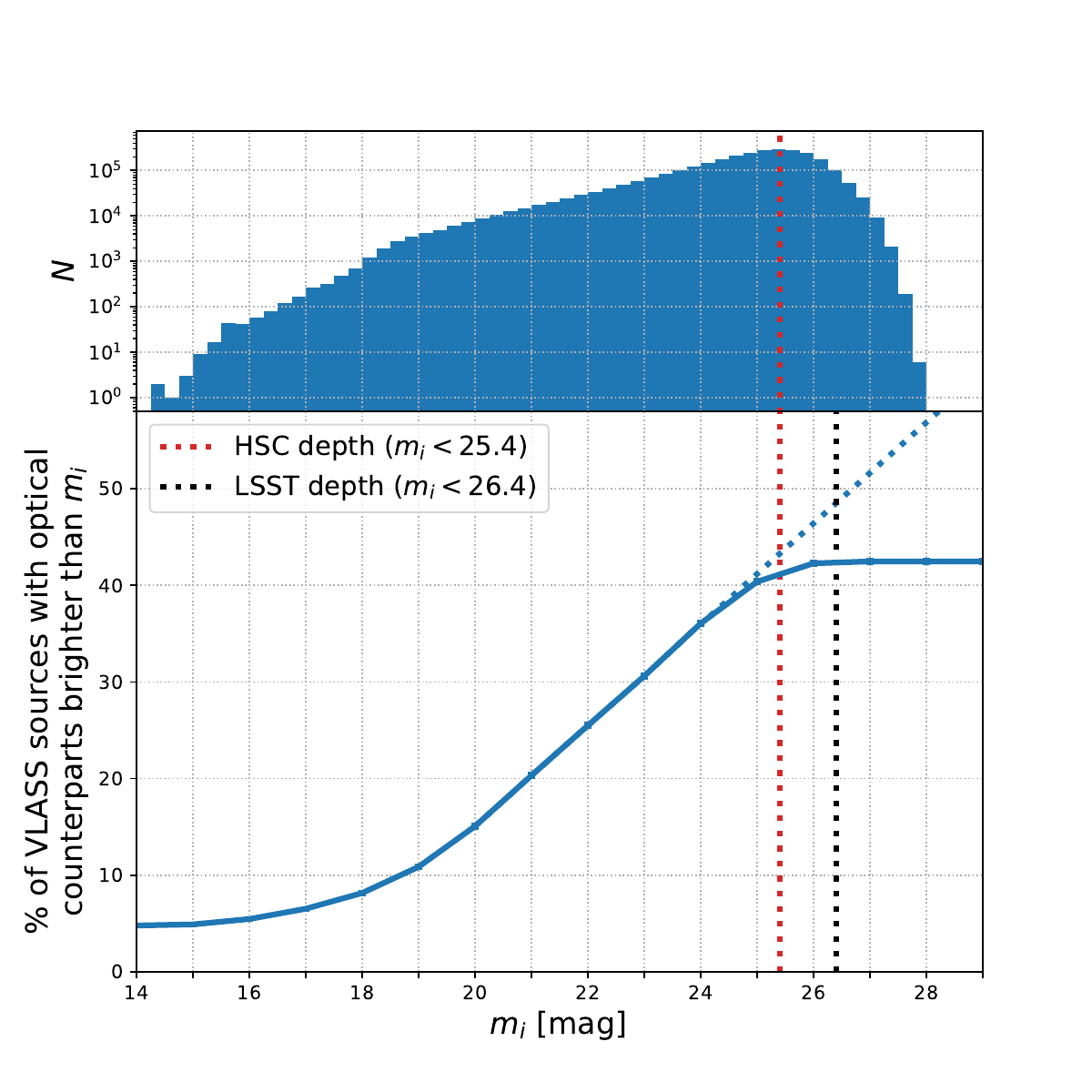}}
    \caption{Panel (a) shows the distribution of angular separations for VLASS/HSC matches (blue line), with the expected and actual matches from random positions shown by the red dotted line and grey histogram respectively.
    The black solid line shows the estimated reliability of a match as a function of separation.
    Panel (b) shows the percentage of VLASS sources with an optical counterpart as a function of $i$-band depth. 
    The upper histogram in panel (b) shows the $i$-band magnitude distribution for HSC.
    }
    \label{figure}
\end{figure}

In Figure \ref{figure}b we show how the fraction of VLASS sources with an HSC counterpart increases with a progressively fainter optical flux threshold.
While the target depth of HSC in the $i$-band is $26.2\,$mag, the survey starts to lose completeness at $i\approx 25.4\,$mag (see Figure \ref{figure}b). 
Using a least squares fit to model the linear increase in fraction of radio sources with a counterpart in the range $20 < m_{i} < 24$, where HSC is dominated by galaxies and still highly complete, we can extrapolate to the target depth of LSST.
Doing this, we predict that $\approx48\,\%$ of VLASS sources should have optical counterparts in LSST.
Given the source density of the VLASS data used here ($55\,\text{deg}^{-2}$) and the $14,000\,\text{deg}^2$ overlap between the LSST and VLASS footprints, 
LSST should be able to identify optical counterparts to $\sim 370,000$ radio sources in each VLASS epoch.

Notably we have only considered one-to-one matches in this analysis.
% It is known that radio galaxies can exhibit complex morphologies, with some structures being detected as multiple radio sources associated with a single optical source.
Radio galaxies can exhibit complex morphologies leading some to appear as multiple sources offset from their optical counterpart.
Only a small fraction ($\lesssim 10\,\%$) of radio sources are expected to exhibit such multi-component morphology \citep{Gordon2023}, so we expect these to have a negligible impact on our `first-order' predictions.
Additionally, the radio flux distributions of sources with and without an HSC counterpart are similar, implying optical depth is more important than radio depth for matching completeness.

A final point worth considering in this Note is the multi-epoch observing strategy of VLASS.
An important aspect of this approach is the ability to combine data from individual epochs to increased the depth of the survey. 
After three epochs, stacking of full quality VLASS images is expected to reach a point source depth of $\approx 350\,\mu\text{Jy}$, detecting $\approx 5\times10^{6}$ sources for a source density of $\approx 145\,\text{deg}^{-2}$.
Interestingly, these additional sources will be at sub-mJy brightnesses, a regime where relatively low redshift star-forming galaxies start to dominate over higher redshift AGN in radio source counts \citep{Condon2012}.
Consequently, the fraction of VLASS sources with a detected optical counterpart is likely to be at least as large in the deep multi-epoch stacked images as for single-epoch observations.
In this scenario, LSST has the potential to observe $\sim 970,000$ optical counterparts to radio sources detected in the VLASS multi-epoch stacks. 
Combining data from VLASS and LSST therefore has the potential to provide a wealth of information on radio sources in the coming years.

%%%%%%%%%%%%%%%%%%%%%%%%%%%%%%%%%%%%%
%%%%%%%%%%%%%%%%%%%%%%%%%%%%%%%%%%%%%

% \begin{acknowledgments}
% A.T. received support from a Hubert Max Thaxton Fellowship and The Ronald E. McNair Postbaccalaureate Achievement Program.
% Y.G. and K.B are supported by National Science Foundation Grant AST 22-06053.
% \end{acknowledgments}

\section*{}\noindent
A.T. received support from a Hubert Max Thaxton Fellowship and The Ronald E. McNair Postbaccalaureate Achievement Program.
Y.G. and K.B are supported by National Science Foundation Grant AST 22-06053.

\vfill
\bibliography{vlass_x_hsc}{}

\begin{thebibliography}{}
\expandafter\ifx\csname natexlab\endcsname\relax\def\natexlab#1{#1}\fi
\providecommand{\url}[1]{\href{#1}{#1}}
\providecommand{\dodoi}[1]{doi:~\href{http://doi.org/#1}{\nolinkurl{#1}}}
\providecommand{\doeprint}[1]{\href{http://ascl.net/#1}{\nolinkurl{http://ascl.net/#1}}}
\providecommand{\doarXiv}[1]{\href{https://arxiv.org/abs/#1}{\nolinkurl{https://arxiv.org/abs/#1}}}

\bibitem[{{Aihara} {et~al.}(2022){Aihara}, {AlSayyad}, {Ando}, {Armstrong},
  {Bosch}, {Egami}, {Furusawa}, {Furusawa}, {Harasawa}, {Harikane}, {Hsieh},
  {Ikeda}, {Ito}, {Iwata}, {Kodama}, {Koike}, {Kokubo}, {Komiyama}, {Li},
  {Liang}, {Lin}, {Lupton}, {Lust}, {MacArthur}, {Mawatari}, {Mineo},
  {Miyatake}, {Miyazaki}, {More}, {Morishima}, {Murayama}, {Nakajima},
  {Nakata}, {Nishizawa}, {Oguri}, {Okabe}, {Okura}, {Ono}, {Osato}, {Ouchi},
  {Pan}, {Plazas Malag{\'o}n}, {Price}, {Reed}, {Rykoff}, {Shibuya},
  {Simunovic}, {Strauss}, {Sugimori}, {Suto}, {Suzuki}, {Takada}, {Takagi},
  {Takata}, {Takita}, {Tanaka}, {Tang}, {Taranu}, {Terai}, {Toba}, {Turner},
  {Uchiyama}, {Vijarnwannaluk}, {Waters}, {Yamada}, {Yamamoto}, \&
  {Yamashita}}]{Aihara2022}
{Aihara}, H., {AlSayyad}, Y., {Ando}, M., {et~al.} 2022, \pasj, 74, 247,
  \dodoi{10.1093/pasj/psab122}

\bibitem[{{Bianco} {et~al.}(2022){Bianco}, {Ivezi{\'c}}, {Jones}, {Graham},
  {Marshall}, {Saha}, {Strauss}, {Yoachim}, {Ribeiro}, {Anguita}, {Bauer},
  {Bauer}, {Bellm}, {Blum}, {Brandt}, {Brough}, {Catelan}, {Clarkson},
  {Connolly}, {Gawiser}, {Gizis}, {Hlo{\v{z}}ek}, {Kaviraj}, {Liu}, {Lochner},
  {Mahabal}, {Mandelbaum}, {McGehee}, {Neilsen}, {Olsen}, {Peiris}, {Rhodes},
  {Richards}, {Ridgway}, {Schwamb}, {Scolnic}, {Shemmer}, {Slater}, {Slosar},
  {Smartt}, {Strader}, {Street}, {Trilling}, {Verma}, {Vivas}, {Wechsler}, \&
  {Willman}}]{Bianco2022}
{Bianco}, F.~B., {Ivezi{\'c}}, {\v{Z}}., {Jones}, R.~L., {et~al.} 2022, \apjs,
  258, 1, \dodoi{10.3847/1538-4365/ac3e72}

\bibitem[{{Condon} {et~al.}(2012){Condon}, {Cotton}, {Fomalont}, {Kellermann},
  {Miller}, {Perley}, {Scott}, {Vernstrom}, \& {Wall}}]{Condon2012}
{Condon}, J.~J., {Cotton}, W.~D., {Fomalont}, E.~B., {et~al.} 2012, \apj, 758,
  23, \dodoi{10.1088/0004-637X/758/1/23}

\bibitem[{{Galvin} {et~al.}(2020){Galvin}, {Huynh}, {Norris}, {Wang},
  {Hopkins}, {Polsterer}, {Ralph}, {O'Brien}, \& {Heald}}]{Galvin2020}
{Galvin}, T.~J., {Huynh}, M.~T., {Norris}, R.~P., {et~al.} 2020, \mnras, 497,
  2730, \dodoi{10.1093/mnras/staa1890}

\bibitem[{{Gordon} {et~al.}(2021){Gordon}, {Boyce}, {O'Dea}, {Rudnick},
  {Andernach}, {Vantyghem}, {Baum}, {Bui}, {Dionyssiou}, {Safi-Harb}, \&
  {Sander}}]{Gordon2021}
{Gordon}, Y.~A., {Boyce}, M.~M., {O'Dea}, C.~P., {et~al.} 2021, \apjs, 255, 30,
  \dodoi{10.3847/1538-4365/ac05c0}

\bibitem[{{Gordon} {et~al.}(2023){Gordon}, {Rudnick}, {Andernach}, {Morabito},
  {O'Dea}, {Achong}, {Baum}, {Bayona-Figueroa}, {Hooper}, {Mingo}, {Morris}, \&
  {Vantyghem}}]{Gordon2023}
{Gordon}, Y.~A., {Rudnick}, L., {Andernach}, H., {et~al.} 2023, \apjs, 267, 37,
  \dodoi{10.3847/1538-4365/acda30}

\bibitem[{{Ivezi{\'c}} {et~al.}(2019){Ivezi{\'c}}, {Kahn}, {Tyson}, {Abel},
  {Acosta}, {Allsman}, {Alonso}, {AlSayyad}, {Anderson}, {Andrew}, {Angel},
  {Angeli}, {Ansari}, {Antilogus}, {Araujo}, {Armstrong}, {Arndt}, {Astier},
  {Aubourg}, {Auza}, {Axelrod}, {Bard}, {Barr}, {Barrau}, {Bartlett}, {Bauer},
  {Bauman}, {Baumont}, {Bechtol}, {Bechtol}, {Becker}, {Becla}, {Beldica},
  {Bellavia}, {Bianco}, {Biswas}, {Blanc}, {Blazek}, {Blandford}, {Bloom},
  {Bogart}, {Bond}, {Booth}, {Borgland}, {Borne}, {Bosch}, {Boutigny},
  {Brackett}, {Bradshaw}, {Brandt}, {Brown}, {Bullock}, {Burchat}, {Burke},
  {Cagnoli}, {Calabrese}, {Callahan}, {Callen}, {Carlin}, {Carlson},
  {Chandrasekharan}, {Charles-Emerson}, {Chesley}, {Cheu}, {Chiang}, {Chiang},
  {Chirino}, {Chow}, {Ciardi}, {Claver}, {Cohen-Tanugi}, {Cockrum}, {Coles},
  {Connolly}, {Cook}, {Cooray}, {Covey}, {Cribbs}, {Cui}, {Cutri}, {Daly},
  {Daniel}, {Daruich}, {Daubard}, {Daues}, {Dawson}, {Delgado}, {Dellapenna},
  {de Peyster}, {de Val-Borro}, {Digel}, {Doherty}, {Dubois},
  {Dubois-Felsmann}, {Durech}, {Economou}, {Eifler}, {Eracleous}, {Emmons},
  {Fausti Neto}, {Ferguson}, {Figueroa}, {Fisher-Levine}, {Focke}, {Foss},
  {Frank}, {Freemon}, {Gangler}, {Gawiser}, {Geary}, {Gee}, {Geha}, {Gessner},
  {Gibson}, {Gilmore}, {Glanzman}, {Glick}, {Goldina}, {Goldstein}, {Goodenow},
  {Graham}, {Gressler}, {Gris}, {Guy}, {Guyonnet}, {Haller}, {Harris},
  {Hascall}, {Haupt}, {Hernandez}, {Herrmann}, {Hileman}, {Hoblitt}, {Hodgson},
  {Hogan}, {Howard}, {Huang}, {Huffer}, {Ingraham}, {Innes}, {Jacoby}, {Jain},
  {Jammes}, {Jee}, {Jenness}, {Jernigan}, {Jevremovi{\'c}}, {Johns}, {Johnson},
  {Johnson}, {Jones}, {Juramy-Gilles}, {Juri{\'c}}, {Kalirai}, {Kallivayalil},
  {Kalmbach}, {Kantor}, {Karst}, {Kasliwal}, {Kelly}, {Kessler}, {Kinnison},
  {Kirkby}, {Knox}, {Kotov}, {Krabbendam}, {Krughoff}, {Kub{\'a}nek},
  {Kuczewski}, {Kulkarni}, {Ku}, {Kurita}, {Lage}, {Lambert}, {Lange},
  {Langton}, {Le Guillou}, {Levine}, {Liang}, {Lim}, {Lintott}, {Long},
  {Lopez}, {Lotz}, {Lupton}, {Lust}, {MacArthur}, {Mahabal}, {Mandelbaum},
  {Markiewicz}, {Marsh}, {Marshall}, {Marshall}, {May}, {McKercher}, {McQueen},
  {Meyers}, {Migliore}, {Miller}, {Mills}, {Miraval}, {Moeyens}, {Moolekamp},
  {Monet}, {Moniez}, {Monkewitz}, {Montgomery}, {Morrison}, {Mueller},
  {Muller}, {Mu{\~n}oz Arancibia}, {Neill}, {Newbry}, {Nief}, {Nomerotski},
  {Nordby}, {O'Connor}, {Oliver}, {Olivier}, {Olsen}, {O'Mullane}, {Ortiz},
  {Osier}, {Owen}, {Pain}, {Palecek}, {Parejko}, {Parsons}, {Pease},
  {Peterson}, {Peterson}, {Petravick}, {Libby Petrick}, {Petry},
  {Pierfederici}, {Pietrowicz}, {Pike}, {Pinto}, {Plante}, {Plate}, {Plutchak},
  {Price}, {Prouza}, {Radeka}, {Rajagopal}, {Rasmussen}, {Regnault}, {Reil},
  {Reiss}, {Reuter}, {Ridgway}, {Riot}, {Ritz}, {Robinson}, {Roby}, {Roodman},
  {Rosing}, {Roucelle}, {Rumore}, {Russo}, {Saha}, {Sassolas}, {Schalk},
  {Schellart}, {Schindler}, {Schmidt}, {Schneider}, {Schneider}, {Schoening},
  {Schumacher}, {Schwamb}, {Sebag}, {Selvy}, {Sembroski}, {Seppala}, {Serio},
  {Serrano}, {Shaw}, {Shipsey}, {Sick}, {Silvestri}, {Slater}, {Smith},
  {Smith}, {Sobhani}, {Soldahl}, {Storrie-Lombardi}, {Stover}, {Strauss},
  {Street}, {Stubbs}, {Sullivan}, {Sweeney}, {Swinbank}, {Szalay}, {Takacs},
  {Tether}, {Thaler}, {Thayer}, {Thomas}, {Thornton}, {Thukral}, {Tice},
  {Trilling}, {Turri}, {Van Berg}, {Vanden Berk}, {Vetter}, {Virieux},
  {Vucina}, {Wahl}, {Walkowicz}, {Walsh}, {Walter}, {Wang}, {Wang}, {Warner},
  {Wiecha}, {Willman}, {Winters}, {Wittman}, {Wolff}, {Wood-Vasey}, {Wu},
  {Xin}, {Yoachim}, \& {Zhan}}]{Ivezic2019}
{Ivezi{\'c}}, {\v{Z}}., {Kahn}, S.~M., {Tyson}, J.~A., {et~al.} 2019, \apj,
  873, 111, \dodoi{10.3847/1538-4357/ab042c}

\bibitem[{{Lacy} {et~al.}(2020){Lacy}, {Baum}, {Chandler}, {Chatterjee},
  {Clarke}, {Deustua}, {English}, {Farnes}, {Gaensler}, {Gugliucci},
  {Hallinan}, {Kent}, {Kimball}, {Law}, {Lazio}, {Marvil}, {Mao}, {Medlin},
  {Mooley}, {Murphy}, {Myers}, {Osten}, {Richards}, {Rosolowsky}, {Rudnick},
  {Schinzel}, {Sivakoff}, {Sjouwerman}, {Taylor}, {White}, {Wrobel},
  {Andernach}, {Beasley}, {Berger}, {Bhatnager}, {Birkinshaw}, {Bower},
  {Brandt}, {Brown}, {Burke-Spolaor}, {Butler}, {Comerford}, {Demorest}, {Fu},
  {Giacintucci}, {Golap}, {G{\"u}th}, {Hales}, {Hiriart}, {Hodge}, {Horesh},
  {Ivezi{\'c}}, {Jarvis}, {Kamble}, {Kassim}, {Liu}, {Loinard}, {Lyons},
  {Masters}, {Mezcua}, {Moellenbrock}, {Mroczkowski}, {Nyland}, {O'Dea},
  {O'Sullivan}, {Peters}, {Radford}, {Rao}, {Robnett}, {Salcido}, {Shen},
  {Sobotka}, {Witz}, {Vaccari}, {van Weeren}, {Vargas}, {Williams}, \&
  {Yoon}}]{Lacy2020}
{Lacy}, M., {Baum}, S.~A., {Chandler}, C.~J., {et~al.} 2020, \pasp, 132,
  035001, \dodoi{10.1088/1538-3873/ab63eb}

\bibitem[{{Villarreal Hern{\'a}ndez} \&
  {Andernach}(2018)}]{VillarrealHernandez2018}
{Villarreal Hern{\'a}ndez}, A.~C., \& {Andernach}, H. 2018, arXiv e-prints,
  arXiv:1808.07178, \dodoi{10.48550/arXiv.1808.07178}

\end{thebibliography}
\bibliographystyle{aasjournal}



\end{document}